\documentclass[notitlepage,aps,12pt,showpacs,showkeys,tightenlines]{revtex4}
\usepackage{amsmath}
\usepackage{amssymb,amsfonts}
\usepackage{bm}
\usepackage[mathcal]{euscript}
\usepackage{graphicx}
\usepackage{psfrag}

\newcommand{\comment}[1]{}

\newcommand{\ie}{\textit{i.e.}}

\newcommand{\mathnotation}[2]{\newcommand{#1}{\ensuremath{#2}}}

\newcommand{\Order}[1]{\ensuremath{\mathcal{O}\!\l(#1\r)}}% Higher order terms.

\renewcommand{\l}{\left}			% \left
\renewcommand{\r}{\right}			% \right
\mathnotation{\pd}{\partial}			% Partial derivative
\mathnotation{\ee}{{\mathrm e}}			% e
\mathnotation{\imi}{\mathrm{i}}			% i
\mathnotation{\ldef}{\mathrel{\raisebox{.069ex}{:}\!\!=}}% Left define
\mathnotation{\rdef}{\mathrel{=\!\!\raisebox{.069ex}{:}}}% Right define
\mathnotation{\dint}{\,{\mathrm{d}}}		% Differential, in an integral

\mathnotation{\iter}{i}				% Iteration
\mathnotation{\xc}{x}				% Spatial coordinate
\mathnotation{\yc}{y}				% Spatial coordinate 2
\mathnotation{\xv}{{\bm{\xc}}}			% Spatial position vector
\mathnotation{\yv}{{\bm{\yc}}}			% Spatial position vector 2

\mathnotation{\Var}{\sigma}			% Variance
\mathnotation{\Rvar}{\sigma}			% Relative variance
\mathnotation{\kv}{{\bm{k}}}			% Wavevector
\mathnotation{\qv}{{\bm{q}}}			% Wavevector 2
\mathnotation{\Qv}{{\bm{Q}}}			% Wavevector 3

\mathnotation{\niterc}{\iter_{1}}		% Onset of superexp. phase
\mathnotation{\niters}{\iter_{2}}		% Onset of exp. phase
\mathnotation{\twotorus}{{\mathcal{T}^2}}	% 2-torus
\mathnotation{\M}{\mathcal{M}}			% Map
\mathnotation{\Mc}{M}				% Linear part of map (comp)
\mathnotation{\ML}{\mathbb{\Mc}}		% Linear part of map (tensor)
\mathnotation{\Heat}{\mathcal{H}}		% Heat operator
\mathnotation{\heat}{h}				% Heat operator kernel
\mathnotation{\eps}{\epsilon}			% Diffusivity

\mathnotation{\BesselJ}{J}

\begin{document}

%\title{Advection--diffusion in a Chaotic Map of the Torus}
\title{Chaotic Mixing in a Torus Map}
\author{Jean-Luc Thiffeault}
\email{jeanluc@mailaps.org}
\affiliation{Department of Applied Physics and Applied Mathematics,
Columbia University, New York, NY 10027}
\altaffiliation[Present address: ]{Department of Mathematics, Imperial College
  London, SW7 2AZ, United Kingdom}
\author{Stephen Childress}
\email{childress@cims.nyu.edu}
\affiliation{Courant Institute of Mathematical Sciences, New York University,
New York, NY 10012}
\date{\today}

\begin{abstract}
The advection and diffusion of a passive scalar is investigated for a map of
the 2-torus.  The map is chaotic, and the limit of almost-uniform stretching
is considered.  This allows an analytic understanding of the transition from a
phase of constant scalar variance (for short times) to exponential decay (for
long times).  This transition is embodied in a short superexponential phase of
decay.  The asymptotic state in the exponential phase is an eigenfunction of
the advection--diffusion operator, in which most of the scalar variance is
concentrated at small scales, even though a large-scale mode sets the decay
rate.  The duration of the superexponential phase is proportional to the
logarithm of the exponential decay rate; if the decay is slow enough then
there is no superexponential phase at all.
\end{abstract}

\maketitle

\textbf{A crucial problem involving fluids in the physical sciences is to
understand the nature of mixing---its efficiency and thoroughness.  Examples
range from the mundane (cream in coffee), to the utilitarian (temperature in
room), the industrial (mixing in chemical reactors), and the planetary (mixing
of ozone in the extratropical stratosphere).  If the flow is not turbulent,
mixing can nevertheless be very efficient, due to a phenomenon called chaotic
advection.  In that case, the flow appears regular, but individual fluid
trajectories are very complicated and lead to a stretching and folding action
that greatly enhances mixing.  Here we discuss mixing for a simple map, and
show that a large-scale, coherent pattern is created that dominates the
diffusive process.}

\section{Introduction}
\label{sec:intro}

It has recently been suggested~\cite{Fereday2002,Wonhas2002} that estimates of
the decay rate of the variance of a passive scalar under the effect of
advection and diffusion~\cite{Antonsen1996,Chertkov1998,Balkovsky1999,Son1999}
do not yield satisfactory results when applied to some simple maps, such as
the inhomogeneous baker's map.~\cite{Farmer1983,Finn1988a,Finn1990} This also
seems to be the case in laboratory experiments on periodic
flows,~\cite{Williams1997,Voth2002} where the decay rate is observed to be
about an order of magnitude slower than the decay rate based on local
arguments, such as the distribution of Lyapunov
exponents.~\cite{VothGollubPrivate} Part of the reason for this is that in
chaotic advection~\cite{Aref1984} (\ie, smooth flows with chaotic Lagrangian
trajectories), far from the highly-turbulent regime, the presence of
slowly-decaying eigenfunctions dominates the long-time decay
rate.~\cite{Pierrehumbert1994,Rothstein1999,Pierrehumbert2000,Fereday2002,%
Wonhas2002} (For the experiments, the presence of regular islands and barriers
is also crucial, but we shall not address this complicated and
poorly-understood issue here.  It suffices to observe that the concentration
field clearly attains an eigenfunction-like regime.~\cite{Rothstein1999}) The
existence of such eigenfunctions of the advection--diffusion operator was
demonstrated convincingly via a numerical approach for the inhomogeneous
baker's map.~\cite{Fereday2002,Wonhas2002} \citet{Sukhatme2002} explained that
the discrepancy is not due to a failure of the local approaches, but because
they assume that the initial scale of variation of the passive scalar is much
smaller than the system size.

Here we propose to use a diffeomorphism of the 2-torus (an extension of
Arnold's cat map~\cite{Arnold}) to further investigate aspects of the decay of
variance and provide some analytical results.  We find that, when the map is
close to uniformly stretching, the decay rate is much faster than indicated by
the distribution of Lyapunov exponents, as was also found in the inhomogeneous
baker's map.~\cite{Fereday2002} In~\citet{Fereday2002} and laboratory
experiments,~\cite{VothGollubPrivate} a slower decay was also observed, but
far from the uniformly-stretching (homogeneous) regime.

The paper is organized as follows.  In Section~\ref{sec:model} we introduce
the map and derive basic expressions for the effect of advection and diffusion
on a passive scalar.  We then analyze the superexponential
(Section~\ref{sec:superexp}) and exponential (Section~\ref{sec:exp}) phases of
diffusion.  The spectrum of variance for the exponential eigenfunction is
derived in Section~\ref{sec:spectrum}, followed by a discussion of the results
in Section~\ref{sec:discussion}.

\section{Advection--diffusion in a Map}
\label{sec:model}

We consider a diffeomorphism of the 2-torus~$\twotorus = [0,1]^2$,
\begin{equation}
  \M(\xv) = \ML\cdot\xv + \bm{\phi}(\xv),
  \label{eq:torusdiffeo}
\end{equation}
where~$\ML$ is a~$2\times2$ nonsingular matrix with integer coefficients
and~$\bm{\phi}(\xv)$ is periodic in both directions with unit period.  We
choose~$\ML$ to have unit determinant, with an eigenvalue larger than one and
the other less than one, so that even in absence of the~$\bm{\phi}$ term~$\M$
is still chaotic.  Specifically, we take
\begin{equation}
  \ML = \begin{pmatrix}2 & 1 \\ 1 & 1 \end{pmatrix};\qquad
  \bm{\phi}(\xv) = \frac{K}{2\pi}
  \begin{pmatrix}\sin2\pi\xc_1 \\ \sin2\pi\xc_1 \end{pmatrix};
  \label{eq:Mphi}
\end{equation}
so that~$\ML\cdot\xv$ is the Arnold cat map and~$\bm{\phi}$ is a wave term
usually associated with the standard map.  The map~$\M$ is area-preserving,
and for~$K=0$ the stretching of phase-space elements is uniform in space.  The
map is always chaotic (the largest Lyapunov exponent is positive).  For
small¬$K$, there are no barriers to transport, such as islands, often
encountered in realistic flows.

We consider the effect of iterating the map and applying the heat operator to
a scalar distribution~$\theta^{(\iter-1)}(\xv)$,
\begin{equation}
  \theta^{(\iter)}(\xv)
  = \Heat_\eps\,\theta^{(\iter-1)}(\M^{-1}(\xv)),
  \label{eq:advdiff}
\end{equation}
where~$\eps$ is the diffusivity, and the heat operator~$\Heat_\eps$ and
kernel~$h_\eps$ are
\begin{equation}
  \Heat_\eps\theta(\xv)
  \ldef \int_{\twotorus}\heat_\eps(\xv-\yv)\,\theta(\yv)\dint\yv;
  \qquad
  h_\eps(\xv) = \sum_{\kv}\exp(2\pi\imi\kv\cdot\xv - \kv^2\eps).
\end{equation}
We Fourier expand~$\theta^{(\iter)}(\xv)$,
\begin{equation}
  \theta^{(\iter)}(\xv) =
  \sum_{\kv}\hat\theta^{(\iter)}_\kv\ee^{2\pi\imi\kv\cdot\xv}
\end{equation}
so that~\eqref{eq:advdiff} becomes
\begin{equation}
  \hat\theta^{(\iter)}(\xv) = 
  \sum_{\qv} \mathbb{T}_{\kv\qv}\,\hat\theta^{(\iter-1)}_\qv,
\end{equation}
with the transfer matrix,
\begin{equation}
  \mathbb{T}_{\kv\qv} \ldef \int_\twotorus
  \exp\l(2\pi\imi\l(\qv\cdot\xv - \kv\cdot\M(\xv)\r) - \eps\,\qv^2\r)
  \dint\xv.
  \label{eq:Tdef}
\end{equation}
We may regard~$\qv$ as the ``initial'' wavenumber, and~$\kv$ as the ``final''
one, with a nonzero~$\mathbb{T}_{\kv\qv}$ denoting a transfer of concentration
from~$\qv$ to~$\kv$ under one application of the map.

For the form of the map given by~\eqref{eq:torusdiffeo} and~\eqref{eq:Mphi},
we have
\begin{equation}
  \mathbb{T}_{\kv\qv} = \ee^{-\eps\,\qv^2} \int_\twotorus
  \exp\l(2\pi\imi\l(\qv - \kv\cdot\ML\r)\cdot\xv
  - \imi\,(k_1 + k_2) K\, \sin(\xc_1)\r)
  \dint\xv.
\end{equation}
The integral in~$\xc_2$ gives a Kronecker delta, and the~$\xc_1$ integral is
readily written as a Bessel function; we thus have
\begin{equation}
  \mathbb{T}_{\kv\qv} = \ee^{-\eps\,\qv^2}\,
  \delta_{0,Q_2}\,\imi^{Q_1}\,
  \BesselJ_{Q_1}\l((k_1+k_2)\,K\r),
  \qquad
  \Qv \ldef \kv\cdot\ML - \qv,
  \label{eq:T}
\end{equation}
where the~$J_Q$ is a Bessel function of the first kind.

\begin{figure}
  \psfrag{variance}{variance, $\Var^{(\iter)}$}
  \psfrag{iteration}{iteration, $\iter$}
  \psfrag{e-15.2i}{$\ee^{-15.2\iter}$}
  \centering\includegraphics[width=.9\columnwidth]{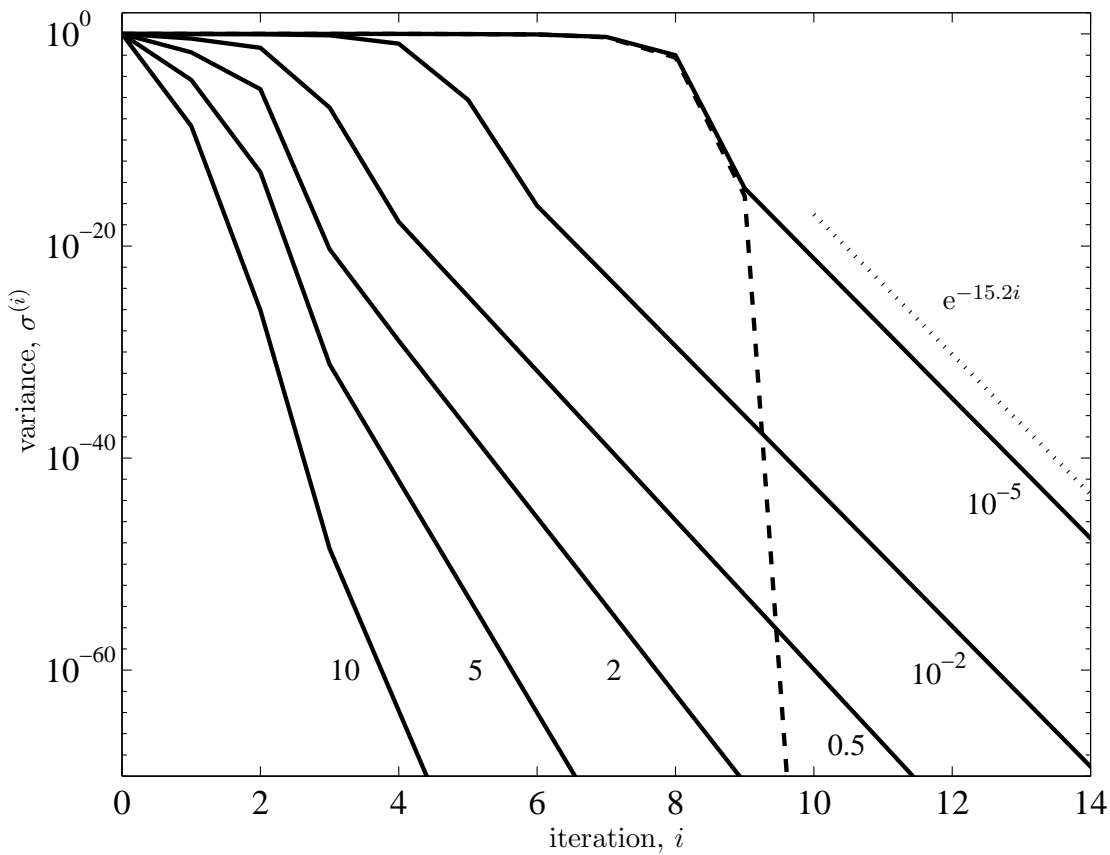}
  \caption{Decay of total variance for varying diffusivity~$\eps$
  and~$K=10^{-3}$.  The onset time of decay is logarithmic in the diffusivity,
  but the asymptotic exponential decay rate becomes independent of the
  diffusivity as~\hbox{$\eps\rightarrow 0$}.  The dashed curve shows the exact
  superexponential solution ($K=0$) for~$\eps=10^{-5}$, and the dotted line is
  the single-mode value from Eq.~\eqref{eq:mudef}.}
  \label{fig:varplot}
\end{figure}

In the absence of diffusion ($\eps=0$), the variance
\begin{equation}
  \Var^{(\iter)} \ldef
  \int_\twotorus \l\lvert\theta^{(\iter)}(\xv)\r\rvert^2 \dint\xv
  = \sum_{\kv} \Var^{(\iter)}_{\kv};
  \qquad
  \Var^{(\iter)}_{\kv} \ldef
  \bigl\lvert\hat\theta^{(\iter)}_{\kv}\bigr\rvert^2,
\end{equation}
is preserved by~\eqref{eq:advdiff} (we assume the spatial mean of~$\theta$ is
zero), and for~$\eps>0$ the variance decays (Fig.~\ref{fig:varplot}).
We consider the case~\hbox{$\eps\ll 1$}, of greatest practical interest.  For
small~$K$, there are three phases: (i) the variance is initially constant (if
the initial scale of variation of the scalar concentration is well above the
diffusive scale, as assumed here); (ii) it then undergoes a rapid
superexponential decay; and (iii) it ultimately decays exponentially at a
fixed rate, independent of~$\eps$, as~$\eps\rightarrow 0$.  In the first
phase, the map has not yet created gradients large enough for the small
diffusion to act.  In the second phase, there is a rapid exponential cascade
to small scales and an associated exponential diffusion, leading to a
superexponential decay.  As the variance is depleted by diffusion, eventually
the system settles into an eigenfunction that sets the exponential decay rate
in the final phase.

The existence of these three phases is
well-known,~\cite{Antonsen1996,Elperin2001,Fereday2002,Wonhas2002,%
Thiffeault2003b} but the exponential phase is the poorly understood, at least
for the case of smooth flows and maps.  We discuss the superexponential phase
briefly in Section~\ref{sec:superexp}, and in Section~\ref{sec:exp} we
describe the exponential phase.  We will see that if the eigenfunction of the
exponential phase decays slowly enough, then there is no superexponential
phase at all.

\section{The Superexponential Phase}
\label{sec:superexp}

Initially, the variance is essentially constant because the tiny diffusivity
can be neglected.  However, there is a cascade of the variance to larger
wavenumbers under the action of~$\M^{-1}$ in~\eqref{eq:advdiff}.  (In this
phase, for small~$K$, we can neglect the $\bm{\phi}$ term
in~\eqref{eq:torusdiffeo}, so that the map~$\M$ is Arnold's cat
map~$\ML\cdot\xv$.)  This is the well-known ``filamentation'' effect in
chaotic flows: the stretching and folding action of the flow causes rapid
variation of the concentration across the folds.  Thus, after a number of
iterations~\hbox{$\niterc \simeq 1 +
(\log\eps^{-1}/\log\Lambda^2)$},\footnote{The extra $1$ in the definition of
$\niterc$ appears because we diffuse at the beginning of the step in
Eq.~\eqref{eq:advdiff}.}  where $\Lambda=(3+\sqrt{5})/2 \simeq 2.618$ is the
largest eigenvalue of~$\ML^{-1}$, the diffusion can no longer be neglected.
For~$\eps=10^{-5}$, we have~\hbox{$\niterc \simeq 6$} (this is always an
overestimate).  We now describe what happens to the variance after diffusion
sets in.
%, for the case of small~$K$.

For small~$K$ and~$\kv$, we have~\hbox{$\BesselJ_{0}\l((k_1+k_2)K\r) \gg
\BesselJ_{1}\l((k_1+k_2)K\r)$}, so we retain only the~$Q_1=0$ term in the
transfer matrix~\eqref{eq:T},
\begin{equation}
  \mathbb{T}_{\kv\qv} = \ee^{-\eps\,\qv^2}\,
  \delta_{\bm{0},\Qv}
  + \Order{(k_1+k_2)^2K^2};
  \label{eq:Tsuperexp}
\end{equation}
Hence, the nonvanishing matrix elements of~$\mathbb{T}$ have~\hbox{$\kv =
\qv\cdot\ML^{-1}$}.
%, and after a few iterations this amounts to~\hbox{$\kv
%  \simeq\Lambda^{i}\,\sdiruv$}, where 
If initially the variance is concentrated in a single wavenumber~$\qv_0$ (\ie,
\hbox{$\Var^{(0)}_{\kv} = 0$} unless \hbox{$\kv=\qv_0$}), then after one
iteration it will all be in~$\qv_0\cdot\ML^{-1}$, after two
in~$\qv_0\cdot\ML^{-2}$, etc.  This amounts to the length of~$\qv$ being
multiplied by a factor~\hbox{$\Lambda > 1$} at each iteration.  But at each
iteration the variance is multiplied by the diffusive decay
factor~$\exp(-2\eps\,\qv^2)$, with~$\qv$ getting exponentially larger.
The total variance is given by
\begin{equation}
  \Var^{(\iter)}
  = \Var^{(0)}\,\exp(-2\eps\,\lVert\qv_0\cdot\ML^{-(\iter-1)}\rVert^2)
  \simeq \Var^{(0)}\,\exp(-2\eps\,\lVert\qv_0\rVert^2\,\Lambda^{2(\iter-1)}),
  \label{eq:Varsuperexp}
\end{equation}
so that the net decay is superexponential.  The superexponential solution is
represented by a dashed line in Fig.~\ref{fig:varplot}, with the solid line
showing the numerical solution for the map~$\M(\xv)$.  The superexponential
solution is valid until about the ninth iteration.\footnote{If, unlike the
present case, the map~$\M$ is not chaotic, that is, it exhibits only algebraic
separation of trajectories, the superexponential stage is replaced by a
faster-than-exponential stage.}  We will revisit this breakdown of the
solution in Section~\ref{sec:exp}.

It is to be noted that a more complicated initial condition also leads to
superexponential decay, albeit with a less well-defined behavior because of
the presence of several modes.  Even an isotropic initial condition can be
expected to have a superxponential phase: the averaging as performed in
\citet{Antonsen1996} is problematic for a cat map in a periodic domain,
because the slope of the stable (contracting) direction is irrational, and yet
the wavevectors are confined to rational slopes.  Hence, for finite~$\kv$ we
cannot expect the averaging to hold.  However, these difficulties are of a
mathematical nature specific to the present problem and do not shed much light
on a more general physical situation.

\section{The Exponential Phase}
\label{sec:exp}

In the superexponential phase we completely neglected the effect of the wave
term in the map~\eqref{eq:torusdiffeo}.  We described the action as a perfect
cascade to large wavenumbers, so that the variance was irrevocably moved to
small scales and dissipated extremely rapidly.  There can be no eigenfunction
in such a situation, since the mode structure changes completely at each
iteration.  This direct cascade process dominates at first, but it is so
efficient that eventually we must examine the effect of the the wave term,
which is felt through the higher-order Bessel functions in the transfer
matrix~\eqref{eq:T}.

Since the long-time exponential decay observed in the numerical results of
Fig.~\ref{fig:varplot} requires the existence of an eigenfunction, we may ask
about the minimum requirement for this.  Clearly if some scalar concentration
is ``left behind'' in a given mode at each iteration, an eigenfunction will
easily form.  The question is then: Is it possible for the scalar
concentration in a given wavenumber to be mapped back onto itself?  This
requires that the diagonal matrix element
\begin{equation}
  \mathbb{T}_{\qv\qv} = \ee^{-\eps\,q_1^2}\,
  \delta_{0,q_1}\,\imi^{q_2}\,
  \BesselJ_{q_2}\l(q_2\,K\r),
  \label{eq:Tdiag}
\end{equation}
be nonzero.  We see from~\eqref{eq:Tdiag} that modes of the form~\hbox{$\qv =
(0\ q_2)$} are mapped to themselves with a nonvanishing amplitude at each
iteration: these are the modes that depend only on the~$\xc_2$ coordinate.
This amplitude vanishes for~$K=0$, since~\hbox{$q_2 \ne 0$} (the~$\qv=0$ mode
is preserved, and of no interest).

\begin{figure}
  \psfrag{log mu2}{$\log\mu^2$}
  \psfrag{K}{$K$}
  \centering\includegraphics[width=.9\columnwidth]{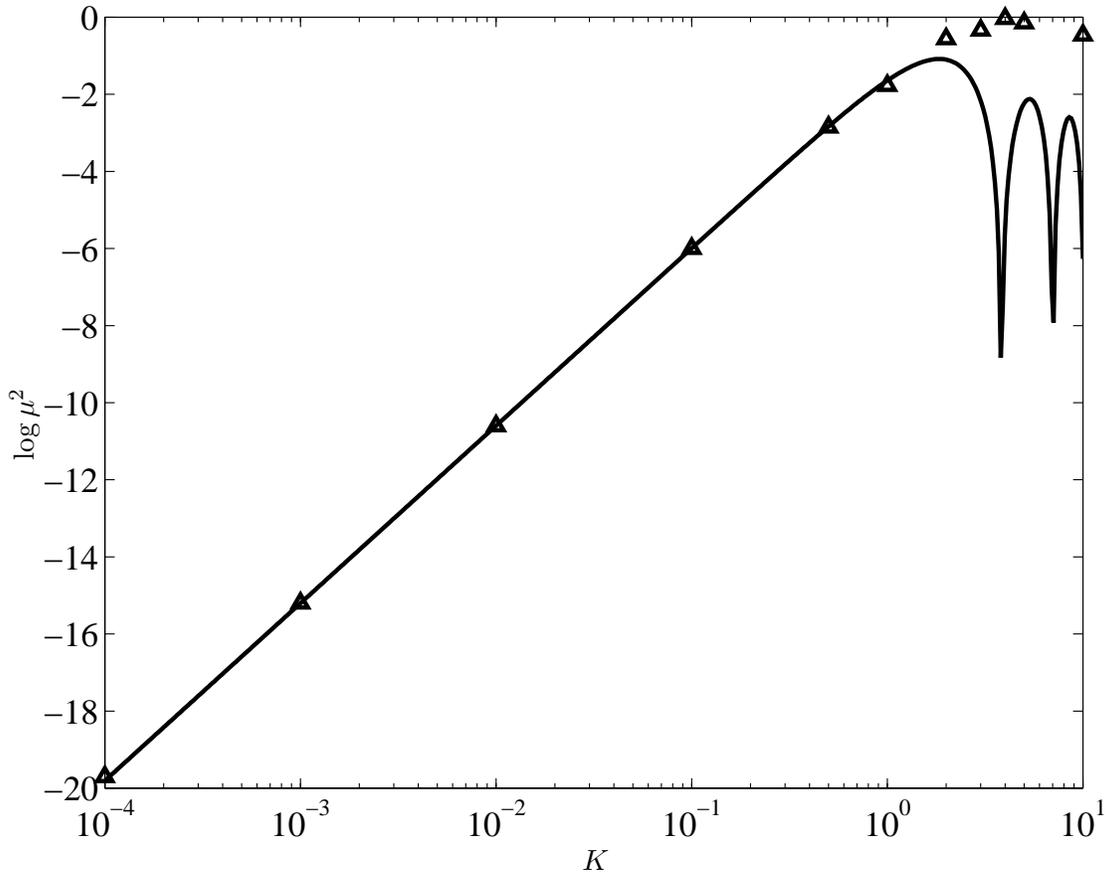}
  \caption{Exponential decay rate~$\log\mu^2$ of the variance
  for~\hbox{$\eps\rightarrow 0$}, as a function of~$K$ and.  The triangles
  denote numerically calculated values, and the solid line is the small-$K$
  expression~\eqref{eq:mudef}.}
  \label{fig:decayrate}
\end{figure}

For small~$K$, the dominant Bessel function after¬$\BesselJ_0$
is~$\BesselJ_1$, so the decay factor~$\mu^2$ for the variance is given by
taking the magnitude of~\eqref{eq:Tdiag},
\begin{equation}
  \mu = \l\lvert\mathbb{T}_{(0\ 1),(0\ 1)}\r\rvert =
  \ee^{-\eps}\,\BesselJ_1\l(K\r) = \tfrac{1}{2}K + \Order{\eps\,K,K^2}.
  \label{eq:mudef}
\end{equation}
Hence, for small~$K$ the decay rate is limited by the~\hbox{$(0\ \ 1)$} mode.
For~\hbox{$\eps\rightarrow 0$}, the decay rate is independent of~$\eps$.
Figure~\ref{fig:decayrate} shows that the single-mode decay rate agrees very
well with the numerical results even for~$K$ close to unity.
In the inhomogeneous baker's map the nearly-superexponential limit is
for~\hbox{$\alpha\rightarrow 1/2$}, where~$\alpha$ is a parameter describing
the inhomogeneity of the map.  For that case the transfer matrix scales in a
manner analogous to here as~\hbox{$\alpha\rightarrow 1/2$}, but many more
modes must be retained due to the presence of discontinuities: all the matrix
coefficients decay as~$(1/2)-\alpha$, with none clearly dominating.  The
single-mode approximation is thus far less accurate.
% See NB XII p. 134.

We can rule out the possibility that the decay is dominated by cycles that
repeat after several iterations (that is,
nonvanishing~$(\mathbb{T}^N)_{\qv\qv}$ for~$N>1$): such cycles must depend on
higher-order Bessel functions that are small compared to~$\BesselJ_1(K)$.
However, as~$K$ is made larger higher-order cycles become dominant and the
situation is much more complicated.

Now that the mechanism of exponential decay is understood (for small~$K$), we
can go back and describe the condition for breakdown of the superexponential
solution discussed at the end of Section~\ref{sec:superexp}.  The
superexponential decay depletes the variance very rapidly until all that is
left is variance in the exponentially decaying mode~\hbox{$\kv_0 \ldef (0\ \
1)$}.  The superexponential phase thus ends when the variance at large
wavenumbers equals that in mode~$\kv_0$.  Assuming that the variance resides
entirely in the~$\kv_0$ mode initially, the condition for this is
\begin{equation}
  \mu^{\niters} =
  \exp(-\eps\,\lVert\kv_0\cdot\ML^{-(\niters-1)}\rVert^2),
  \label{eq:super2exp}
\end{equation}
where~$\mu$ is the decay factor of the variance in the~$\kv_0$ mode, given by
Eq.~\eqref{eq:mudef}, and the right-hand side is the superexponential
solution~\eqref{eq:Varsuperexp}.  After
substituting~\hbox{$\l\lVert\kv_0\cdot\ML^{-(\niters-1)}\r\rVert \simeq
\Lambda^{\niters-1}$}, Eq.~\eqref{eq:super2exp} must be solved numerically
for~$\niters$: for~$K=10^{-3}$ and~$\eps=10^{-5}$, we
have~\hbox{$\niters\simeq9.2$}.  This is in fine agreement with the transition
from superexponential to exponential in Fig.~\ref{fig:varplot}.

If~$\eps\ll 1$, Eq.~\eqref{eq:super2exp} has the approximate solution
\begin{equation}
  \niters \simeq 1 + \frac{\log\l(\eps^{-1}\log\mu^{-1}\r)}{\log\Lambda^2},
  \label{eq:super2expapprox}
\end{equation}
which gives~\hbox{$\niters\simeq8$} for~$K=10^{-3}$,~$\eps=10^{-5}$.
Subtracting~$\niterc = 1+\log\eps^{-1}/\log\Lambda^2$, the onset of the
superexponential phase (Section~\ref{sec:superexp}), we find that the duration
of the superexponential phase is roughly
\begin{equation}
  \niters - \niterc \simeq \frac{\log\log\mu^{-1}}{\log\Lambda^2},
  \label{eq:phase2duration}
\end{equation}
which is independent of~$\eps$ (at leading order), and has a weak dependence
on the decay rate~$\log\mu$.  Unless~$\mu$ is very small (recall
that~\hbox{$0<\mu<1$}), the superexponential phase is very short.  In fact,
for~\hbox{$\log\mu^{-1} < 1$} the decay of the~$(0\ \ 1)$ mode is slow
enough that there is no superexponential phase at all, as indicated by the
negative right-hand side in~\eqref{eq:phase2duration}.  We can thus speculate
that it is unlikely that the superexponential phase can be observed in
experiments, since there~$\mu$ tends to be close to unity.  \comment{Get value
from Greg and Jerry.}

Note that $\eps$ has to be extremely small for~\eqref{eq:phase2duration} to
hold: for~$K=10^{-3}$,~$\eps=10^{-5}$,~\eqref{eq:phase2duration}
gives~\hbox{$\niters - \niterc \simeq 1$}, whereas the unapproximated
(numerical) result is~\hbox{$\niters - \niterc \simeq 2.2$}.  The error
on~\eqref{eq:super2expapprox} and~\eqref{eq:phase2duration} scales
as~$\log\log\eps^{-1}$.
% See superexp2exp.nb.

\section{Variance Spectrum of the Eigenfunction}
\label{sec:spectrum}

The long-wavelength mode discussed in Section~\ref{sec:exp} is the bottleneck
that determines the decay rate (for small~$K$).  But this mode does
not dominate the structure of the eigenfunction.  In fact, a very small
amount of the total variance actually resides in that bottleneck mode: the
variance is concentrated at small scales.  We now derive the variance spectrum
of the eigenfunction.

The variance is taken out of the~\hbox{$(0\ \ 1)$} mode in the same manner as
described in Section~\ref{sec:superexp}: there is a cascade from that mode to
larger wavenumbers through the action of~$\ML^{-1}$.  Neglecting the~$K$ term,
the cascade proceeds from~$\kv_0=(0\ \ 1)$ to~$\kv_1$, $\kv_2$, $\kv_3$, \dots
etc., as
\begin{equation}
  (0\ \ 1) \rightarrow (-1\ \ 2) \rightarrow (-3\ \ 5)
  \rightarrow (-8\ \ 13) \rightarrow \ldots.
  \label{eq:cascade}
\end{equation}
The~$\kv_n$ become more and more aligned with the stable (contracting)
direction of the map as we proceed down the cascade.  The amplitude of the
wavenumber is multiplied at each step by a
factor~\hbox{$\Lambda=(3+\sqrt{5})/2 \simeq 2.618$}, the largest eigenvalue
of~$\ML^{-1}$.

\begin{figure}
  \psfrag{mu2}{{\large$\mu^2$}}
  \psfrag{nu0}{{\large$\nu_0$}}
  \psfrag{nu1}{{\large$\nu_1$}}
  \psfrag{nu2}{{\large$\nu_2$}}
  \psfrag{iteration i}{{iteration $i$}}
  \psfrag{iteration i-1}{{iteration $i-1$}}
  \psfrag{k0}{{\large $\kv_0$}}
  \psfrag{k1}{{\large $\kv_1$}}
  \psfrag{k2}{{\large $\kv_2$}}
  \psfrag{k3}{{\large $\kv_3$}}
  \centering\includegraphics[width=.6\columnwidth]{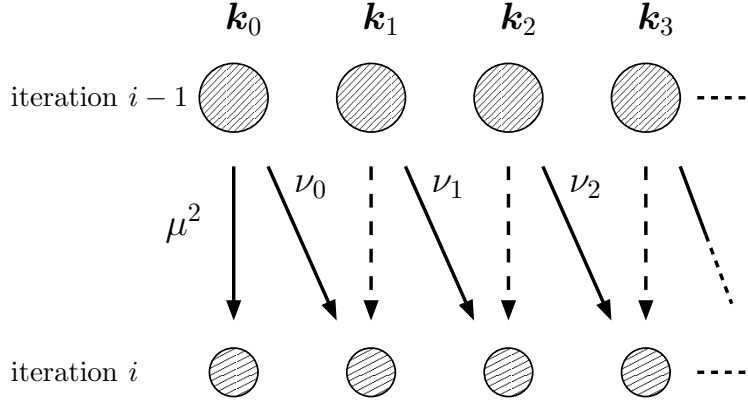}
  %\centering\includegraphics[width=\columnwidth]{figs/modecasc.eps}
  \caption{Schematic representation of the cascade of variance for an
  eigenfunction.  The solid arrows represent a direct transfer of
  concentration, the dashed an ``effective'' transfer of amplitude~$\mu^2$ due
  to the eigenfunction property.  In our aproximation, only the~$\kv_0$ mode
  has a direct transfer of concentration to itself.}
  \label{fig:modecasc}
\end{figure}

The exponential decay rate suggests that the scalar concentration is in an
eigenfunction of the advection--diffusion operator.  Assuming this to be the
case, Fig.~\ref{fig:modecasc} illustrates the transfer of variance between
modes for an iteration of the map.
At each iteration, the eigenfunction property implies that the variance in
each wavenumber is decreased by a uniform factor~\hbox{$\mu^2 < 1$}.  This is
illustrated by the vertical arrows in Fig.~\ref{fig:modecasc}.  The dashed
arrows do not represent a direct transfer of variance, since for small~$K$
only the variance in the~$\kv_0$ mode is actually mapped back onto itself
after one iteration (this is denoted by a solid vertical arrow).  Rather,
there is an \emph{effective} (indirect) transfer occurring because of the
cascade~\eqref{eq:cascade}: most of the variance in each mode is mapped to the
next one down the cascade following the diagonal arrows in
Fig.~\ref{fig:modecasc}.

The decrease in variance for each of the diagonal arrows is diffusive and is
given by the factor~\hbox{$\nu_n = \exp(-2\eps\,\kv_{n}^2)$}.  If we denote
by~\hbox{$\Var_{\kv_n}^{(i)} \ldef \lvert\hat\theta_{\kv_n}^{(i)}\rvert^2$}
the variance in mode~$\kv_n$ at the~$i$th iteration, we have
\begin{subequations}
\begin{alignat}{2}
  \Var_{\kv_n}^{(i)}
  &= \mu^2\,\Var_{\kv_n}^{(i-1)},&\qquad &n=0,1,\ldots,\\
  \Var_{\kv_n}^{(i)}
  &= \nu_{n-1}\,\Var_{\kv_{n-1}}^{(i-1)},&\qquad &n=1,2,\ldots.
  \label{eq:sigmarecur}
\end{alignat}
These two recurrences can be combined to give
\end{subequations}
\begin{equation}
  \Rvar^{(i)}(\kv_n) = \frac{\nu_{n-1}\,\nu_{n-2}\,\cdots\nu_0}{\mu^{2n}}
  = \mu^{-2n}\,\exp\biggl(-2\eps\sum_{m=0}^{n-1}\kv_m^2\biggr),
  \label{eq:Rvardef}
\end{equation}
where the \emph{relative variance} in the~$n$th mode is defined
as~\hbox{$\Rvar^{(i)}(\kv_n) \ldef
\Rvar_{\kv_n}^{(i)}/\Rvar_{\kv_0}^{(i)}$}.  The magnitude of the wavenumber
is given by the exponential recursion,
\begin{equation}
  \l\lVert\kv_n\r\rVert \simeq \Lambda\l\lVert\kv_{n-1}\r\rVert
  \quad\Longrightarrow\quad
  \l\lVert\kv_n\r\rVert \simeq \Lambda^{n}\l\lVert\kv_0\r\rVert
  = \Lambda^{n}\,,
  \label{eq:krecur}
\end{equation}
which allows us to solve for~$n$,
\begin{equation}
  n = \log\l\lVert\kv_n\r\rVert/\log\Lambda
\end{equation}
and rewrite~\eqref{eq:Rvardef} as
\begin{equation}
  \Rvar^{(i)}(\kv_n) \simeq \l\lVert\kv_n\r\rVert^{-2\log\mu/\log\Lambda}
  \exp\l(-2\eps\,\kv_n^2/\Lambda^2\r),
  \label{eq:Rvar2}
\end{equation}
where we retained only the~$\kv_{n-1}^2$ term of the sum in~\eqref{eq:Rvardef}
and used~\eqref{eq:krecur}.  The right-hand side of Eq.~\eqref{eq:Rvar2} for
the relative variance does not (and should not if we really have an
eigenfunction) depend on the iteration number, $i$, and depends only on~$n$
through~$\kv_n$.  We thus let~\hbox{$\kv_n$} be a continuous variable~$\kv$,
with~\hbox{$\kv = k\,(\cos\theta,\sin\theta)$}, and
drop~$\iter$; from Eq.~\eqref{eq:Rvar2} we then have
\begin{equation}
  \Rvar(k,\theta) = \widetilde{\Rvar}(k)\,k^{-1}\,\delta(\theta-\theta_0),
  \label{eq:fullspectrum}
\end{equation}
the spectrum of relative variance, with
\begin{equation}
  \widetilde{\Rvar}(k) = k^{2\zeta}\,\exp\l(-2\eps\, k^2/\Lambda^2\r),
  \qquad
  \zeta \ldef -\log\mu/\log\Lambda.
  \label{eq:spectrum}
\end{equation}
The factor~$k^{-1}\,\delta(\theta-\theta_0)$ in~\eqref{eq:fullspectrum}
reflects the alignment of the vectors~$\kv$ with the stable (contracting)
direction of~$\ML$, which is at an angle~$\theta_0$.  We thus have essentially
a one-dimensional spectrum.  The spectrum function then satisfies
\begin{equation}
  \widetilde{\Rvar}(k) = \int\Rvar(k,\theta)\,k\dint\theta.
\end{equation}

\begin{figure}
  \psfrag{k}{$k$}
  \psfrag{relative variance}{relative variance, $\widetilde{\Rvar}(k)$}
  \centering\includegraphics[width=.9\columnwidth]{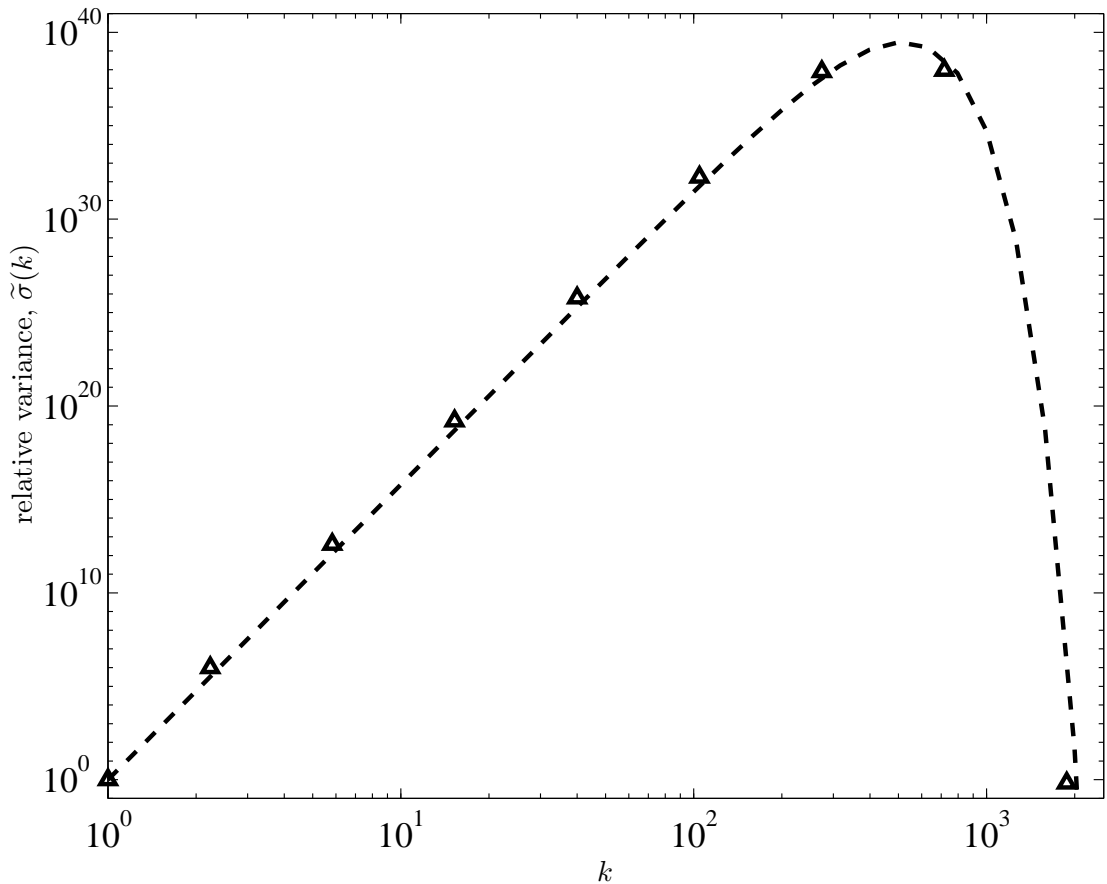}
  \caption{Spectrum function of relative variance after~$12$ iterations
  for~$K=10^{-3}$, $\eps=10^{-4}$.  The dashed line is the theoretical curve
  given by~\eqref{eq:spectrum}, and the triangles are numerical results.}
  \label{fig:spectrum}
\end{figure}

The spectrum function~\eqref{eq:spectrum} is plotted in Fig.~\ref{fig:spectrum}
and compared with numerical results for small~$K$, showing excellent
agreement.  Since~\hbox{$\mu^2 < 1$} and~\hbox{$\Lambda > 1$}, we conclude
that~\hbox{$\zeta > 0$} always.  This implies that there is more variance at
the large wavenumbers than at the slowest-decaying mode~$\kv_0$.  The slope of
the spectrum~$\widetilde{\Rvar}(k)$ is considerably shallower than the
Batchelor~$k^{-1}$ spectrum,~\cite{Batchelor1959} consistent with the results
for the baker's map.~\cite{Fereday2002}  This reflects the extreme efficiency
of the cascade, a consequence of the nearly-uniform stretching, in that small
scales are generated with great ease and the spectrum is therefore skewed
towards large wavenumbers.

To know just how much more variance is at the large wavenumbers, we find the
maximum of~\eqref{eq:spectrum}, which is at
\begin{equation}
  k_{\mathrm{m}} = \Lambda\,\l(\zeta/2\eps\r)^{1/2},
  \qquad
  \Rvar(k_{\mathrm{m}})
  = k_{\mathrm{m}}^{2\zeta}\,\ee^{-\zeta}
  = k_{\mathrm{m}}^{2\zeta}\,\mu^{\log\Lambda}.
  \label{eq:kmax}
\end{equation}
The peak wavenumber thus scales as~$\eps^{-1/2}$, the same scaling as the
dissipation scale.  From~\eqref{eq:kmax}, the relative variance in that peak
wavenumber scales as~$\eps^{-\zeta}$.  The wavenumber~$k_{\mathrm{m}}$ gives
an indication of the largest wavenumber that must be included in a numerical
calculation to capture the decay of variance correctly.  However, if the
truncation size is smaller, the decay rate in the exponential phase is still
captured properly, since it is determined by the~\hbox{$(0\ \ 1)$} mode.

\section{Discussion}
\label{sec:discussion}

We summarize the three phases of chaotic mixing in smooth flows for the case
of small diffusivity.  In the first phase the variance is approximately
conserved, and the chaotic flow (or map) creates large gradients in the scalar
concentration through its stretching and folding action.  This is usually
called the \emph{stirring} phase.  In the second phase, the variance (that is,
the squared-amplitude of each mode with the total mean subtracted) starts to
decrease superexponentially, because the exponential cascade to small scales
is compounded by the exponential efficiency of diffusion
(Section~\ref{sec:superexp}).  This is the first of two \emph{mixing} phases
(superexponential and exponential), where diffusion plays an important role.
This superexponential phase might not occur if the exponential decay rate of
the slowest-decaying eigenfunction is slow enough.  For very small
diffusivity, the duration of the superexponential phase is independent of
diffusivity.

Unless the stretching is completely uniform, the superexponential phase comes
to an end because though it rapidly depletes any variance contained in the
small scales, some is left behind because of dispersion.  What remains is the
eigenfunction of the advection--diffusion operator with the largest eigenvalue
(all eigenvalues have modulus less than one), which then decays exponentially.
The decay rate of this eigenfunction is determined by its slowest-decaying
part, in the present case the~\hbox{$(0 \ \ 1)$} mode (Section~\ref{sec:exp}).
The structure (spectrum) of this eigenmode is readily described as a balance
between the eigenfunction property (modes are mapped to themselves with
uniform amplitude) and a cascade to large wavenumbers
(Section~\ref{sec:spectrum}).  In the present case of a map with nearly
uniform stretching, the spectrum of the eigenfunction has most of its variance
concentrated at large wavenumbers, even though the small wavenumber
mode~\hbox{$(0 \ \ 1)$} dictates the rate of decay.

The decay rate of variance is outrageously fast in a map so close to being
superexponential.  Nevertheless, the manner in which the asymptotic regime is
attained and the possibility of analytic results provide insight into the
formation of the eigenfunction through the interplay of the slowest-decaying
mode and the cascade to large wavenumbers.  As $K$ is made larger, the decay
rates are more reasonable and a remnant of the mechanism presented here still
applies.

The decay rate in the present case is completely unrelated to the Lyapunov
exponent or its distribution.  For small~$K$, the distribution of the Lyapunov
exponent is peaked at~$\log\Lambda$ and has a very narrow standard deviation.
\comment{Of order $K$, $K^{1/2}$, what?}  But here the asymptotic exponential
decay rate is of order~$\log K$, so the decay becomes faster
as~\hbox{$K\rightarrow 0$}.  This is due to the system being close to the
uniform stretching (cat map) limit, which is unlikely to be the case in
physical situations.  Any theory based on the distribution Lyapunov exponents
cannot in this case predict the decay rate, since a \emph{global} mode
dominates.  For the theory of \citet{Antonsen1996}, there is the further
problem that, as in \citet{Fereday2002}, averaging over angles is not possible
here, since for small~$K$ the stable manifold and the gradient of the initial
condition have a nearly constant angle with respect to each other.  If the
initial condition itself is taken as isotropic, then the irrationality of the
slope of the contracting direction becomes problematic
(Section~\ref{sec:superexp}).

\citet{Sukhatme2002} point out that what determines the regime of decay (\ie,
eigenfunction or local) is the scale of the initial scalar concentration.  In
our case, as we make that initial scale smaller we find the same asymptotic
decay rate.  This is due to a weak dispersion (due to the wave) from the large
to the small wavenumbers which allows the system to develop its preferred
(slowest-decaying) large-scale eigenfunction.  The only way to get a faster
rate is to completely remove the slowest-decaying eigenfunction from the
initial condition, which never happens in practice.  A smaller initial scale
of variation does however lead to faster overall decay because its effect is
to lengthen the initial superexponential scale.  This is because the weak
dispersion needs time to build the eigenfunction to an amplitude where it can
rise above the other ambient modes.

The large-scale eigenfunctions can lead not only to faster decay but also
slower (as in~\citet{Fereday2002}), when compared to local, Lyapunov-exponent
based approaches.~\cite{Antonsen1996,Balkovsky1999,Son1999} In both cases, it
is the highly-ordered nature of the system (due to the large-scale, coherent
nature of the initial scalar field and flow, but also to periodic boundary
conditions and walls) that gives the discrepancy.  We also observe a slower
decay for larger~$K$, but no analytical theory has yet been developed to
adequately describe that regime.

We observe numerically that as~$K$ is made large the spectrum of variance
tends to concentrate in small wavenumbers, possibly due to the presence of a
strong dispersion competing with the direct cascade to small
scales.~\cite{STF}  In that limit the cascade to large wavenumbers is no
longer described by the linear part~$\ML$ of the map, so there is no clear
separation between the eigenfunction property and the cascade.  An
investigation of the decay rate and spectrum in this large~$K$, wave-dominated
limit will be the subject of future work.

\begin{acknowledgments}
The authors thank A. H. Boozer, J. P. Gollub, G. A. Voth, and D. Lazanja for
helpful discussions, and two anonymous referees for very constructive
comments.  J.-L.T. was supported by the National Science Foundation and the
Department of Energy under a Partnership in Basic Plasma Science grant,
No.~DE-FG02-97ER54441.
\end{acknowledgments}

%\bibliography{books,journals_abbrev,chaotic_advection,VothGollub}

\begin{thebibliography}{22}
\expandafter\ifx\csname natexlab\endcsname\relax\def\natexlab#1{#1}\fi
\expandafter\ifx\csname bibnamefont\endcsname\relax
  \def\bibnamefont#1{#1}\fi
\expandafter\ifx\csname bibfnamefont\endcsname\relax
  \def\bibfnamefont#1{#1}\fi
\expandafter\ifx\csname citenamefont\endcsname\relax
  \def\citenamefont#1{#1}\fi
\expandafter\ifx\csname url\endcsname\relax
  \def\url#1{\texttt{#1}}\fi
\expandafter\ifx\csname urlprefix\endcsname\relax\def\urlprefix{URL }\fi
\providecommand{\bibinfo}[2]{#2}
\providecommand{\eprint}[2][]{\url{#2}}

\bibitem[{\citenamefont{Fereday et~al.}(2002)\citenamefont{Fereday, Haynes,
  Wonhas, and Vassilicos}}]{Fereday2002}
\bibinfo{author}{\bibfnamefont{D.~R.} \bibnamefont{Fereday}},
  \bibinfo{author}{\bibfnamefont{P.~H.} \bibnamefont{Haynes}},
  \bibinfo{author}{\bibfnamefont{A.}~\bibnamefont{Wonhas}}, \bibnamefont{and}
  \bibinfo{author}{\bibfnamefont{J.~C.} \bibnamefont{Vassilicos}},
  \bibinfo{journal}{Phys. Rev. E} \textbf{\bibinfo{volume}{65}},
  \bibinfo{pages}{035301(R)} (\bibinfo{year}{2002}).

\bibitem[{\citenamefont{Wonhas and Vassilicos}(2002)}]{Wonhas2002}
\bibinfo{author}{\bibfnamefont{A.}~\bibnamefont{Wonhas}} \bibnamefont{and}
  \bibinfo{author}{\bibfnamefont{J.~C.} \bibnamefont{Vassilicos}},
  \bibinfo{journal}{Phys. Rev. E} \textbf{\bibinfo{volume}{66}},
  \bibinfo{pages}{051205} (\bibinfo{year}{2002}).

\bibitem[{\citenamefont{{Antonsen, Jr.} et~al.}(1996)\citenamefont{{Antonsen,
  Jr.}, Fan, Ott, and Garcia-Lopez}}]{Antonsen1996}
\bibinfo{author}{\bibfnamefont{T.~M.} \bibnamefont{{Antonsen, Jr.}}},
  \bibinfo{author}{\bibfnamefont{Z.}~\bibnamefont{Fan}},
  \bibinfo{author}{\bibfnamefont{E.}~\bibnamefont{Ott}}, \bibnamefont{and}
  \bibinfo{author}{\bibfnamefont{E.}~\bibnamefont{Garcia-Lopez}},
  \bibinfo{journal}{Phys. Fluids} \textbf{\bibinfo{volume}{8}},
  \bibinfo{pages}{3094} (\bibinfo{year}{1996}).

\bibitem[{\citenamefont{Balkovsky and Fouxon}(1999)}]{Balkovsky1999}
\bibinfo{author}{\bibfnamefont{E.}~\bibnamefont{Balkovsky}} \bibnamefont{and}
  \bibinfo{author}{\bibfnamefont{A.}~\bibnamefont{Fouxon}},
  \bibinfo{journal}{Phys. Rev. E} \textbf{\bibinfo{volume}{60}},
  \bibinfo{pages}{4164} (\bibinfo{year}{1999}).

\bibitem[{\citenamefont{Chertkov et~al.}(1998)\citenamefont{Chertkov,
  Falkovich, and Kolokolov}}]{Chertkov1998}
\bibinfo{author}{\bibfnamefont{M.}~\bibnamefont{Chertkov}},
  \bibinfo{author}{\bibfnamefont{G.}~\bibnamefont{Falkovich}},
  \bibnamefont{and}
  \bibinfo{author}{\bibfnamefont{I.}~\bibnamefont{Kolokolov}},
  \bibinfo{journal}{Phys. Rev. Lett.} \textbf{\bibinfo{volume}{80}},
  \bibinfo{pages}{2121} (\bibinfo{year}{1998}).

\bibitem[{\citenamefont{Son}(1999)}]{Son1999}
\bibinfo{author}{\bibfnamefont{D.~T.} \bibnamefont{Son}},
  \bibinfo{journal}{Phys. Rev. E} \textbf{\bibinfo{volume}{59}},
  \bibinfo{pages}{R3811} (\bibinfo{year}{1999}).

\bibitem[{\citenamefont{Farmer et~al.}(1983)\citenamefont{Farmer, Ott, and
  Yorke}}]{Farmer1983}
\bibinfo{author}{\bibfnamefont{J.~D.} \bibnamefont{Farmer}},
  \bibinfo{author}{\bibfnamefont{E.}~\bibnamefont{Ott}}, \bibnamefont{and}
  \bibinfo{author}{\bibfnamefont{J.~A.} \bibnamefont{Yorke}},
  \bibinfo{journal}{Physica D} \textbf{\bibinfo{volume}{7}},
  \bibinfo{pages}{153} (\bibinfo{year}{1983}).

\bibitem[{\citenamefont{Finn and Ott}(1988)}]{Finn1988a}
\bibinfo{author}{\bibfnamefont{J.~M.} \bibnamefont{Finn}} \bibnamefont{and}
  \bibinfo{author}{\bibfnamefont{E.}~\bibnamefont{Ott}},
  \bibinfo{journal}{Phys. Rev. Lett.} \textbf{\bibinfo{volume}{60}},
  \bibinfo{pages}{760} (\bibinfo{year}{1988}).

\bibitem[{\citenamefont{Finn and Ott}(1990)}]{Finn1990}
\bibinfo{author}{\bibfnamefont{J.~M.} \bibnamefont{Finn}} \bibnamefont{and}
  \bibinfo{author}{\bibfnamefont{E.}~\bibnamefont{Ott}},
  \bibinfo{journal}{Phys. Fluids B} \textbf{\bibinfo{volume}{2}},
  \bibinfo{pages}{916} (\bibinfo{year}{1990}).

\bibitem[{\citenamefont{Williams et~al.}(1997)\citenamefont{Williams, Marteau,
  and Gollub}}]{Williams1997}
\bibinfo{author}{\bibfnamefont{B.}~\bibnamefont{Williams}},
  \bibinfo{author}{\bibfnamefont{D.}~\bibnamefont{Marteau}}, \bibnamefont{and}
  \bibinfo{author}{\bibfnamefont{J.~P.} \bibnamefont{Gollub}},
  \bibinfo{journal}{Phys. Fluids} \textbf{\bibinfo{volume}{9}},
  \bibinfo{pages}{2061} (\bibinfo{year}{1997}).

\bibitem[{\citenamefont{Voth et~al.}(2002)\citenamefont{Voth, Haller, and
  Gollub}}]{Voth2002}
\bibinfo{author}{\bibfnamefont{G.~A.} \bibnamefont{Voth}},
  \bibinfo{author}{\bibfnamefont{G.}~\bibnamefont{Haller}}, \bibnamefont{and}
  \bibinfo{author}{\bibfnamefont{J.~P.} \bibnamefont{Gollub}},
  \bibinfo{journal}{Phys. Rev. Lett.} \textbf{\bibinfo{volume}{88}},
  \bibinfo{pages}{254501} (\bibinfo{year}{2002}).

\bibitem[{\citenamefont{Voth and Gollub}()}]{VothGollubPrivate}
\bibinfo{author}{\bibfnamefont{G. A.}~\bibnamefont{Voth}} \bibnamefont{and}
  \bibinfo{author}{\bibfnamefont{J. P.}~\bibnamefont{Gollub}},
  \bibinfo{howpublished}{private communication}.

\bibitem[{\citenamefont{Aref}(1984)}]{Aref1984}
\bibinfo{author}{\bibfnamefont{H.}~\bibnamefont{Aref}}, \bibinfo{journal}{J.
  Fluid Mech.} \textbf{\bibinfo{volume}{143}}, \bibinfo{pages}{1}
  (\bibinfo{year}{1984}).

\bibitem[{\citenamefont{Rothstein et~al.}(1999)\citenamefont{Rothstein, Henry,
  and Gollub}}]{Rothstein1999}
\bibinfo{author}{\bibfnamefont{D.}~\bibnamefont{Rothstein}},
  \bibinfo{author}{\bibfnamefont{E.}~\bibnamefont{Henry}}, \bibnamefont{and}
  \bibinfo{author}{\bibfnamefont{J.~P.} \bibnamefont{Gollub}},
  \bibinfo{journal}{Nature} \textbf{\bibinfo{volume}{401}},
  \bibinfo{pages}{770} (\bibinfo{year}{1999}).

\bibitem[{\citenamefont{Pierrehumbert}(2000)}]{Pierrehumbert2000}
\bibinfo{author}{\bibfnamefont{R.~T.} \bibnamefont{Pierrehumbert}},
  \bibinfo{journal}{Chaos} \textbf{\bibinfo{volume}{10}}, \bibinfo{pages}{61}
  (\bibinfo{year}{2000}).

\bibitem[{\citenamefont{Pierrehumbert}(1994)}]{Pierrehumbert1994}
\bibinfo{author}{\bibfnamefont{R.~T.} \bibnamefont{Pierrehumbert}},
  \bibinfo{journal}{Chaos Solitons Fractals} \textbf{\bibinfo{volume}{4}},
  \bibinfo{pages}{1091} (\bibinfo{year}{1994}).

\bibitem[{\citenamefont{Sukhatme and Pierrehumbert}(2002)}]{Sukhatme2002}
\bibinfo{author}{\bibfnamefont{J.}~\bibnamefont{Sukhatme}} \bibnamefont{and}
  \bibinfo{author}{\bibfnamefont{R.~T.} \bibnamefont{Pierrehumbert}},
  \bibinfo{journal}{Phys. Rev. E} \textbf{\bibinfo{volume}{66}},
  \bibinfo{pages}{056032} (\bibinfo{year}{2002}).

\bibitem[{\citenamefont{Arnold}(1989)}]{Arnold}
\bibinfo{author}{\bibfnamefont{V.~I.} \bibnamefont{Arnold}},
  \emph{\bibinfo{title}{Mathematical Methods of Classical Mechanics}}
  (\bibinfo{publisher}{Springer-Verlag}, \bibinfo{address}{New York},
  \bibinfo{year}{1989}), \bibinfo{edition}{2nd} ed.

\bibitem[{\citenamefont{Thiffeault}(2003)}]{Thiffeault2003b}
\bibinfo{author}{\bibfnamefont{J.-L.} \bibnamefont{Thiffeault}},
  \bibinfo{journal}{Phys. Lett. A}  (\bibinfo{year}{2003}), \bibinfo{note}{in
  press}, \eprint{arXiv:nlin.CD/0105026}.

\bibitem[{\citenamefont{Elperin et~al.}(2001)\citenamefont{Elperin, Kleeorin,
  Rogachevskii, and Sokoloff}}]{Elperin2001}
\bibinfo{author}{\bibfnamefont{T.}~\bibnamefont{Elperin}},
  \bibinfo{author}{\bibfnamefont{N.}~\bibnamefont{Kleeorin}},
  \bibinfo{author}{\bibfnamefont{I.}~\bibnamefont{Rogachevskii}},
  \bibnamefont{and} \bibinfo{author}{\bibfnamefont{D.}~\bibnamefont{Sokoloff}},
  \bibinfo{journal}{Phys. Rev. E} \textbf{\bibinfo{volume}{63}},
  \bibinfo{pages}{046305} (\bibinfo{year}{2001}).

\bibitem[{\citenamefont{Batchelor}(1959)}]{Batchelor1959}
\bibinfo{author}{\bibfnamefont{G.~K.} \bibnamefont{Batchelor}},
  \bibinfo{journal}{J. Fluid Mech.} \textbf{\bibinfo{volume}{5}},
  \bibinfo{pages}{113} (\bibinfo{year}{1959}).

\bibitem[{\citenamefont{Childress and Gilbert}(1995)}]{STF}
\bibinfo{author}{\bibfnamefont{S.}~\bibnamefont{Childress}} \bibnamefont{and}
  \bibinfo{author}{\bibfnamefont{A.~D.} \bibnamefont{Gilbert}},
  \emph{\bibinfo{title}{Stretch, Twist, Fold: The Fast Dynamo}}
  (\bibinfo{publisher}{Springer-Verlag}, \bibinfo{address}{Berlin},
  \bibinfo{year}{1995}).

\end{thebibliography}

\end{document}